\documentclass[12pt]{article}
\usepackage{amsfonts,amsmath,amssymb}
\usepackage{amssymb}

\begin{document}

\begin{center}
\vskip 3cm {\Large\bf On phase transition signal in inelastic
collision}

\vskip 1cm J.Manjavidze\\

\vskip 0.5cm JINR, Dubna, Russia\footnote{ On leave in absence:
Andronikashvili Institute of Physics, Tbilisi, Georgia}
\end{center}

\begin{abstract}
The paper is devoted to retrieval of the first order phase
transition signal in the inelastic collisions. The primary intent
is to show that the experimentally observable signal exist iff the
multiplicity is sufficiently large. We discuss corresponding
phenomenology from the point of view of experiment.

\end{abstract}


{\bf I}\setcounter{equation}{0}

The question of possibility to observe first order phase
transition in the hadron and heavy ion collisions is discussed
widely at present time \cite{1}. The aim of paper is discuss the
phenomenology of that problem.

The first order phase transition in statistics appears in the
result of creation of local critical fluctuations (e.g. bubbles of
vapor if the (liquid$\to$gas) transition is considered), in
contrast to the second order phase transition where the whole
system undergo the transition. The dimension of such fluctuations
increase, if they are under-critical, and the whole system in
result undergo the transition.

It is hard to imagine that exactly such picture appears in the
hadron or ion inelastic collisions. The point is that in the
event-by-event measurement the dimension of bubbles, {\it if they
exist}, may be both smaller or larger critical one. Therefore, in
the best case, we observe the mixture of two-phase medium and the
question how one may increase on experiment the weight of events
with under-critical bubbles is the first problem. It is evident
that the density, usually used in statistics, can not be
introduced as the "order parameter" since only the result of
particles $production$ process is observed.

In the paper \cite{vhmw} the "chemical potential", $\mu(n,s)$, was
offered as the "order parameter". It is the work which is
necessary for creation of one particle and it was shown that if
the role of under-critical babbles dominate then $\mu(n,s)$ must
decrease with number of produced particles $n$. In another words,
it was offered to define the boiling, i.e. creation of
under-critical bubbles, through $intensity$ of the process of
particles production (evaporation).

This is the main idea why the very high multiplicity (VHM)
processes were considered. It is not important what additional
criterium is used searching first order phase transition. In any
case we must consider the VHM domain to have intensive production
of particles. In addition, the kinetic degrees of freedom must be
suppressed in the VHM region.

Therefore, the special attention will be given to the VHM
processes. Then exist approximation \cite{vhmw}: \begin{equation}
\mu(n,s)\simeq-\frac{T(n,s)}{n} \ln\sigma_n(s)
\label{1.1}\end{equation} in this multiplicity region. Here $T$ is
the mean energy, including mass, of produced particles, i.e. $T$
is associated with temperature, and $\sigma_n$ is the normalized
to unite multiple production cross section which can be considered
in the VHM region as the "partition function" of the $equilibrium$
system. The equilibrium condition will be defined later, see
inequality (\ref{b}), definition (\ref{22}) and \cite{physrep}
where the detailed explanation was given. See also the footnote 5.

Continuing the analogy with thermodynamics one can say that
$(-T\ln\sigma_n)/n$ is the Gibbs free energy per one particle.
Then $\mu$ can be interpreted as the "chemical potential" measured
with help of $observed$ free particles\footnote{Notice that one
may consider $n$ as the multiplicity in the experimentally
observable range of phase space.}.

The definition (\ref{1.1}) is quiet general. It can be used both
for hadron-hadron and ion-ion collisions, both for low and high
energies, both for "boiling" media of coloured partons and
colorless hadrons. Definition (\ref{1.1}) is model free and
operates only with "external" directly measurable parameters. The
single indispensable condition: we work in the VHM region.

We will discuss in the paper the chance of experimental
measurement of $\mu(n,s)$ defined by (\ref{1.1}), what kind
uncertainties hides it from $experimental$ point of view noting
the the cross sections in VHM domain are small. The correction to
(\ref{1.1}) are not essential in the VHM region but nevertheless
the field-theoretical definition of chemical potential in using
Wigner functions formalism \cite{carr} will be published, see also
\cite{physrep}.

{\bf II}.

It is necessary to remind main steps toward (\ref{1.1}) to
understand hidden phenomenological uncertainties. The starting
point \cite{vhmw} was the generating function \begin{equation}
\rho(z,s)= \sum_{n=0}^\infty z^n\sigma_n(s),~
\rho(1,s)=1,~\sigma_n=0~{\rm
at}~n>n_{max}=\sqrt{s}/m,\label{}\end{equation} where $m$ is the
hadron mass. One may use inverse Mellin transformation:
\begin{equation} \sigma_n(s)=\frac{1}{2\pi i}\oint
\frac{dz}{z^{n+1}}\rho(z,s)\label{mel}\end{equation} to find
$\sigma_n$ if $\rho(z,s)$ is known. One may calculate integral
(\ref{mel}) by saddle point method. The equation (of state):
\begin{equation} n=z\frac{\partial}{\partial z}\ln\rho(z,s)\label{eq1}\end{equation} defines
mostly essential value $z=z(n,s)$. Therefore, only
\begin{equation} z<z_{max}=z(n_{max},s)\label{}\end{equation} have the physical
meaning.

One may write $\rho(z,s)$ in the form: \begin{equation}
\rho(z,s)=\exp\left\{ \sum_{l=0}^\infty z^l
b_l(s)\right\},\label{2.4}\end{equation} where the "Mayer group
coefficient" $b_l$ can be expressed through correlators $c_k(s)$:
$$ b_l(s)=\sum_{k=l}^\infty \frac{(-1)^{(k-l)}}{l!(k-l)!}c_k(s).$$
Let us assume now that we have Poisson distribution, i.e. if in
the sum: \begin{equation} \ln\rho_(z,s)=\sum_k \frac{(z-1)^k}
{k!}c_k(s)\label{}\end{equation} one may leave first term, then it
is easy to see that \begin{equation}
z(n,s)=n/c_1(s),~c_1(s)\equiv\bar{n}(s), \label{1.8}\end{equation}
are essential and in the VHM region: \begin{equation}
\ln\sigma_n(s)=-n\ln\frac{n}{c_1(s)}(1+O(1/\ln n))=-n\ln
z(n,s)(1+O(1/\ln n)).\label{esti}\end{equation} Therefore, in
considered case with $c_k=0,~k>1,$ exist following asymptotic
estimation for $n>>1$:
\begin{equation} \ln\sigma_n\simeq-n\ln z(n,s),\label{1.10}\end{equation} i.e. $\sigma_n$
is defined in VHM region mainly by the solution of Eq.(\ref{eq1})
and the correction can not change this conclusion. It will be
shown that that kind estimation is hold for arbitrary asymptotics
of $\sigma_n$. The definition (\ref{1.1}) based on this
observation.

If we understand $\sigma_n$ as the "partition function" in the VHM
region then $z$ is the $activity$ usually introduced in
statistical physics. Correspondingly the chemical potential $\mu$
is defined trough $z$: \begin{equation} \mu=T\ln z.
\label{mu}\end{equation} Combining this definition with estimation
(\ref{1.10}) we define $\sigma_n$ through $\mu$. But, $if$ this
estimation does not depend from asymptotics of $\sigma_n$ over
$n$, i.e. if it has general meaning, then it can be used for
definition of $\mu(n,s)$ through $\sigma_n(s)$ and $T(n,s)$ at
$n>>1$. Just this formal idea is realized in (\ref{1.1}): it can
be shown in Sec.III that (\ref{1.1}) is correct at the
asymptotical value of $n$.

{\bf III.}

Now we will make the important step. To put in a good order our
intuition it is useful to consider $\rho(z,s)$ as the $nontrivial$
function of $z$. In statistical physics the thermodynamical limit
is considered for this purpose. In our case the finiteness of
energy $\sqrt{s}$ and of the hadron mass $m$ put obstacles on this
way since the system of produced particles necessarily belongs to
the energy-momentum surface\footnote{It must be noted that the
canonical thermodynamic system belongs to the energy-momentum
shell because of the energy exchange, i.e. interaction, with
thermostat. The width of the shell is defined by the temperature.
But in particle physics there is no thermostat and the physical
system completely belongs to the energy momentum surface.}. But we
can continue $theoretically$ $\sigma_n$ to the range $n>n_{max}$
and consider $\rho(z,s)$ as the nontrivial function of $z$. This
step hides the assumption that nothing new appear at $n>n_{max}$,
i.e. the VHM interval $\bar{n}<<n<n_{max}$ is sufficiently wide to
represent main physical processes.

Let us consider the analog generating function which has the first
$n<n_{max}$ coefficient of expansion over $z$ equal to $\sigma_n$
and higher coefficients for $n\geq n_{max}$ are deduced from
continuation of theoretical value of $\sigma_n$ to $n\geq
n_{max}$. Then the inverse Mellin transformation (\ref{mel}) gives
a good estimation of $\sigma_n$ through this generating function
if the fluctuations near $z(n,s)$ are Gaussian or, it is the same,
if
\begin{equation} \left.\frac{|2n-z^3\partial^3\ln\rho(z,s)/\partial
z^3|}{|n+z^2\partial^2 \ln\rho(z,s)/\partial
z^2|^{3/2}}\right|_{z=z(n,s)}<<1. \label{z}\end{equation} Notice
that if the estimation (\ref{1.10}) is generally rightful then one
can easily find that l.h.s. of (\ref{z}) is $\sim1/n^{1/2}$.
Therefore, one may consider $\rho(z,s)$ as the nontrivial function
of $z$ considering $z(n,s)<z_{max}$ if $\bar{n}<<n<n_{max}$.

Then it is easily deduce that the asymptotics of $\sigma_n(s)$ is
defined by the leftmost singularity, $z_c$, of function
$\rho(z,s)$ since, as it follows from Eq.(\ref{eq1}), the
singularity "attracts" the solution $z(n,s)$ {\it in the VHM
region}. In result we may classify asymptotics of $\sigma_n$ in
the VHM region if (\ref{z}) is hold.

Our problem is reduced to the definition of possible location of
leftmost singularity of $\rho(z,s)$ over $z>0$\footnote{The
singularities in complex $z$ plane will not be considered since
they lead only to oscillations in multiplicity distribution.}. It
must be stressed that the character of singularity is not
important for definition of $\mu(n,s)$ in the VHM region at least
with $O(1/\ln n)$ accuracy. One may consider only three
possibility at $n\to\infty$: (I) $z(n,s)\to z_c=1$; (II)
$z(n,s)\to z_c,~~1<z_c<\infty$; (III) $z(n,s)\to z_c=\infty$. The
structure of complex $z$ plane is much more complicate but for our
purpose the above described picture is sufficient.

Correspondingly one may consider only three type of asymptotics in
the VHM region: (I) $\sigma_n>O(e^{-n})$; (II)
$\sigma_n=O(e^{-n})$. Such asymptotics is typical for hard
processes with large transverse momenta, like for jets
\cite{physrep}; (III) $\sigma_n<O(e^{-n})$. That asymptotic
behavior is typical for multiperipheral-like kinematics, where the
longitudinal momenta of produced particles are noticeably higher
than the transverse ones \cite{physrep}.

Therefore the case (I) is the best candidate for phase transition
since in this case the cross sections are comparatively large in
the VHM region, i.e. particles "intensively" produced in that
case. Notice that if (I) is not realized in nature then the (II)
kind processes would dominate in the VHM region.

Let us consider now the estimation (\ref{1.1}). It follows from
(\ref{mel}) that, up to the preexponential factor,
\begin{equation} \ln\sigma(n,s)\approx-n\ln z(n,s)+
\ln\rho(z(n,s),s).\label{4.2}\end{equation} We want to show that,
in a vide range of $n$ from VHM region,
\begin{equation} n\ln z(n,s)\sim \ln\rho(z(n,s),s).\label{}\end{equation}

Let as consider now the mostly characteristic examples.

(I) {\it Singularity at $z=1$}. The physical meaning of
singularity at $z=1$ may be illustrated by the droplet model
\cite{lee}. The Mayer's group coefficient, see (\ref{2.4}), for
cluster from $l$ particle is $ b_l(\beta)\sim \exp\{-\beta\tau
l^{(d-1)/d}\},$ where $\tau l^{(d-1)/d}$, $l>>1$, is the surface
tension energy, $d$ is the dimension. Therefore, if $d>1$ the
series over $l$ in (\ref{2.4}) diverges at $z=1$.

This case was considered in \cite{vhmw} in details. In the used
lattice gas approximation  $\ln{z}(n) \sim n^{-5}$ and
$\ln\sigma_n\approx -n^{-4}=-n\ln z(n)(1+O(1/n)).$ Notice that the
simplest droplet model predicts unphysical asymptotics:
$\sigma_n\to const$ in the VHM region.

(II) {\it Singularity at $1<z_c<\infty$}. Let us consider one jet
contribution: $\ln\rho(z,s)=-\gamma\ln(1-\bar{n}_j(s)(z-1))$. In
this case $z(n,s)=z_c(1-\gamma/n),~n>>\gamma,$ and
$\ln\sigma_n=-n\ln z(n,s)(1+O(\ln n/n)).$

(III) {\it Singularity at $z=\infty$}. For $k$ Pomeron exchange:
$\ln\rho(z,s)= c_k(s)(z-1)^k$. In this case $z(n)=
(n/kc_k)^{1/k}>>1$ and $\ln\sigma_n\approx -n\ln z(n)(1+O(1/\ln
n)).$

One can conclude:

(i) The definition (\ref{1.1}) in the VHM region is rightful since
the correction falls down with $n$. On this stage we can give only
the qualitative estimation of corrections. Nevertheless
(\ref{1.1}) gives the correct $n$ dependence in the VHM region.

(ii) Activity $z(n,s)$ tends to $z_c$ from the right in the case
(I) and from the left if we have the case (II) or (III).

(iii) The accuracy of estimation of the chemical potential
(\ref{1.1}) increase from (III) to (I).

{\bf IV.}

The temperature $T$ is the next problem. The temperature is
introduced usually using Kubo-Martin-Schwinger (KMS) periodic
boundary conditions. But this way assumes from the very beginning
that the system (a) is equilibrium and (b) is surrounded by
thermostat through which the temperature is determined. The first
condition (a) we take as the simplification which gives the
equilibrium state.

The second one (b) is the problem since there is no thermostat in
particle physics. For this reason we introduce the temperature as
the Lagrange multiplier $\beta=1/T$ of energy conservation law
\cite{physrep}. In such approach the condition that the system is
in equilibrium with thermostat replaced by the condition that the
fluctuations in vicinity of $\beta$ are Gaussian.

The interesting for us $\rho(z,{s})$ we define through inverse
Laplace transform of $\rho(z,\beta)$: \begin{equation}
\rho(z,s)=\int \frac{d\beta}{2\pi i\sqrt{s}} e^{\beta\sqrt{s}}
\rho(z,\beta). \label{}\end{equation} It is known that if the
interaction radii is finite, i.e. the hadron mass is finite, then
the equation (of state): \begin{equation}
\sqrt{s}=-\frac{\partial}{\partial\beta}\rho(z,\beta)
\label{eq}\end{equation} have real positive solution $\beta(n,s)$
at $z=z(n,s)$. We will assume that the fluctuations near
$\beta(n,{s})$ are Gaussian. This means that the inequality
\cite{physrep}:
\begin{equation} \left.\frac{|\partial^3\ln \rho(z,\beta)/\partial\beta^3|}
{|\partial^2\ln\rho(z,\beta)/\partial\beta^2|^{3/2}}
\right|_{z=z(n,s),\beta=\beta(n,s)}<<1 \label{b}\end{equation} is
satisfied. Therefore, we prepare the formalism to find
"thermodynamic" description of the processes of particle
production assuming that this $S$-matrix condition of equilibrium
(\ref{b}) is hold\footnote{Introduction of $\beta(n,s)$ allows to
describe the system of large number of degrees of freedom in terms
of single parameter $\beta(n,s)$, i.e. it is nothing but the
useful trick. It is no way for this reason to identify entirely
$1/\beta(n,s)$ with thermodynamic temperature where it has the
self-contained physical sense. Nevertheless path-integral
representation of $\rho(\beta,z)$ defined from $S$-matrix
coincides with Feynman-Kac representation of grand partition
function \cite{physrep} if (\ref{b}) is hold. It must be noted
also that the energy spectrum of produced particle in this case
have Boltzmann form, $e^{-\beta\eta}$.}.

I want to underline that our thermal equilibrium condition
(\ref{b}) have absolute meaning: if it is not satisfied then
$\beta(n,s)$ loses every sense since the expansion in vicinity of
$\beta(n,s)$ leads to the asymptotic series. In this case only the
dynamical description of $S$-matrix can be used.

It is not hard to see \cite{physrep} that \begin{equation}
\frac{\partial^l}{\partial\beta^l}\ln R(z,\beta)
|_{z=z(n,s),\beta=\beta(n,s)}=<\prod_{i=1}^l(\eta_i-<\eta>)>_{n,s}
\label{22}\end{equation} is the $l$-point energy correlator, where
$<...>_{n,s}$ means averaging over all events with given
multiplicity and energy. Therefore (\ref{b}) means "relaxation of
$l$-point correlations", $l>2$, measured in units of the
dispersion of energy fluctuations, $l=2$. One can note here the
difference of our definition of thermal equilibrium from
thermodynamical one \cite{bogol}.

{\bf V.}

We may conclude that:

(i) The definition of chemical potential (\ref{1.1}) was
discussed. This important observable can be measured on the
experiment directly. Chemical potential, $\mu(n,s)$, must decrease
in the VHM region if the first order phase transition occur, case
(I), and it rise in opposite case, see (II) and (III), see
Sec.III.

(ii) We are forced to assume that the energy and the multiplicity
are sufficiently large, i.e. the experimental value $z^{exp}(n,s)$
is sufficiently close to $z_c=1$. In opposite case the leading
leftmost singularity over $z$ would not be "seen" on experiment
and the production processes constitutes from the complicated
mixture of subprocesses.

(iii) The cross section $\sigma_n$ falls down rapidly with $n$ and
for this reason the VHM events are hardly observable. One may
avoid this problem considering the finite energy heavy ion
collisions as the most candidates of processes described by
methods of thermodynamics and $z_c$ is easier "reachable" in this
case.

(iv) One can define $z(n,s)$ also directly from Eq.(\ref{eq1}):
\begin{equation} n=z\frac{\partial}{\partial z}\ln\sum_n
z^n\sigma_n^{exp}(s),\label{22a}\end{equation} using experimental
values $\sigma_n^{exp}(s)$. But comparing (\ref{22a}) with
definition (\ref{1.10}), \begin{equation} n\ln
z\simeq-\ln\sigma_n^{exp}(s), \label{23}\end{equation} it seems
that last one gives more definite value of $z^{exp}(n,s)$ than the
"integral" equation (\ref{22a}) especially since the statistical
errors are large in the VHM region and the theoretical correction
to Eq.(\ref{23}) are small, $\sim1/n$.

Summarizing the results we conclude: {\it if the energy is
sufficiently large, i.e. if $z_{max}$ is sufficiently close to
$z_c=1$, if the multiplicity is sufficiently large, so that
(\ref{b}) is satisfied and $z(n,s)$ can be sufficiently close to
$z_c$, then one may have confident answer on the question:
observable or not the first order phase transition in hadron/ion
collisions. The heavy ion collisions are favorable to observe the
phase transition.}

\vskip 0.5cm {\bf Acknowledgements.}

I would like to thank participants of 7-th International Workshop
on the "Very High Multiplicity Physics" (JINR, Dubna) for
stimulating discussions. I am grateful to V.Priezzhev,
A.Sissakian, A.Sorin and V.Kekelidze for valuable attention.


\end{document}